\documentclass{ws-ijmpd}
\begin{document}

\markboth{A.A. Ivanov, S.V. Matarkin and L.V. Timofeev}
{Reconstruction of Cherenkov radiation signals from EAS}

\title{Reconstruction of Cherenkov radiation signals from extensive air showers of cosmic rays using data of a wide field-of-view telescope}

\author{A.A.~Ivanov*, S.V.~Matarkin and L.V.~Timofeev}

\address{Shafer Institute for Cosmophysical Research \& Aeronomy,
Yakutsk 677980, Russia\\
*ivanov@ikfia.ysn.ru}

\maketitle

\begin{abstract}
The operation of a wide field-of-view (WFOV) Cherenkov telescope is described. The detection of extensive air showers (EAS) of cosmic rays (CR) is based upon the coincidence with signals from the Yakutsk array. The data acquisition system of the telescope yields signals connected with EAS development parameters: presumably, shower age and position of shower maximum in the atmosphere. Here we describe the method of signal processing used to reconstruct Cherenkov radiation signals induced by CR showers.
An analysis of signal parameters results in the confirmation of the known correlation of the duration of the Cherenkov radiation signal with the distance to the shower core.
The measured core distance dependence is used to set an upper limit to the dimensions of the area along the EAS axis where the Cherenkov radiation intensity is above half-peak amplitude.

\end{abstract}

\keywords{Cosmic rays, extensive air showers, Cherenkov radiation}

\section{Introduction}
\label{sec:Introduction}
Ultra-high energy cosmic rays (UHECR) entering Earth's atmosphere create a cascade of secondary particles. In this cascade there are myriad charged particles moving at speeds greater than $c/n$, where $c$ is the velocity of light in vacuum and $n$ is the index of refraction in air. These particles emit coherent Cherenkov radiation~\cite{Tamm} that potentially contains information about the energy, composition and arrival direction of the primary particle that initiated the extensive air shower (EAS)\cite{Dyak}.

Since the first observation by Galbraith and Jelley~\cite{Jelley} and a systematic measurement of air Cherenkov radiation properties in the Pamir experiment~\cite{Chudakov}, a number of EAS arrays have been equipped with Cherenkov radiation detectors. Particularly, in the Yakutsk array experiment these detectors are used to estimate the energy and mass composition of the primaries~\cite{Dyak,JETP2007}.

In a majority of previous measurements, analog signal readout systems were used that had narrow bandwidth, which restricted the possibility of pulse-shape reconstruction of the Cherenkov radiation from EAS, or detectors were designed for measurement of the integral signal~\cite{Dyak,Turver,Blanca}.
Recently, digital data acquisition system was implemented in the Tunka-133 Cherenkov array consisting of a set of photomultiplier tubes (PMT)~\cite{Tunka}.
In this paper, we describe a method used for reconstructing the pulse shape of the Cherenkov radiation from the EAS as detected using a WFOV Cherenkov telescope (hereinafter `telescope') based on the coincidence of signals with the Yakutsk array.
The aim is to reconstruct shower characteristics.

\section{The Yakutsk Array Experiment}
\label{sec:Experiment}
\begin{figure}[t]
\begin{minipage}{0.54\textwidth}
\resizebox{1.01\textwidth}{!}{\includegraphics{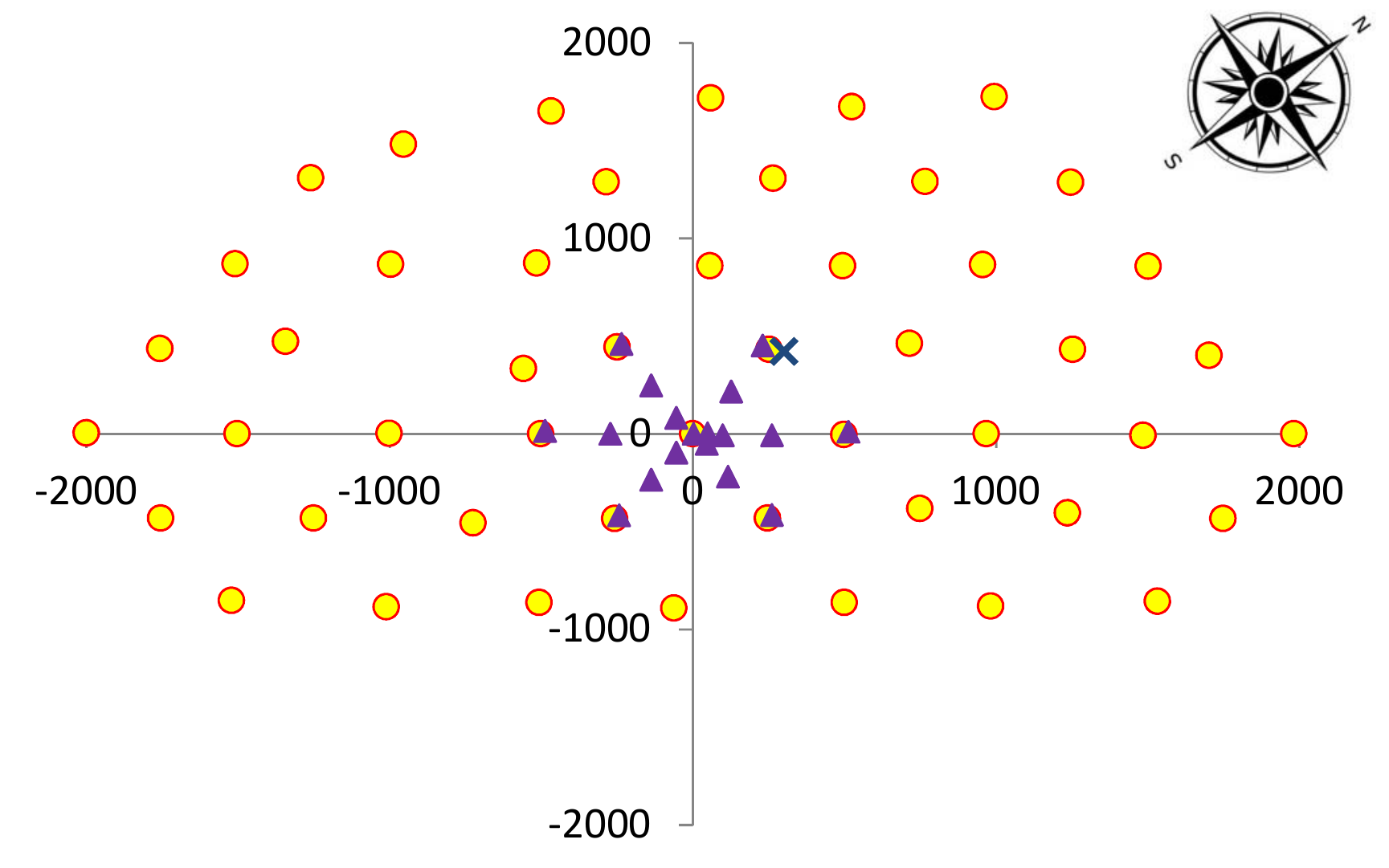}}
\end{minipage}
\begin{minipage}{0.45\textwidth}
\resizebox{0.7\textwidth}{!}{\includegraphics{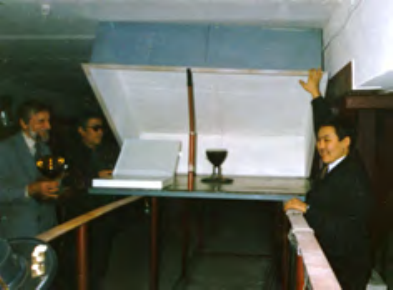}}
\resizebox{0.7\textwidth}{!}{\includegraphics{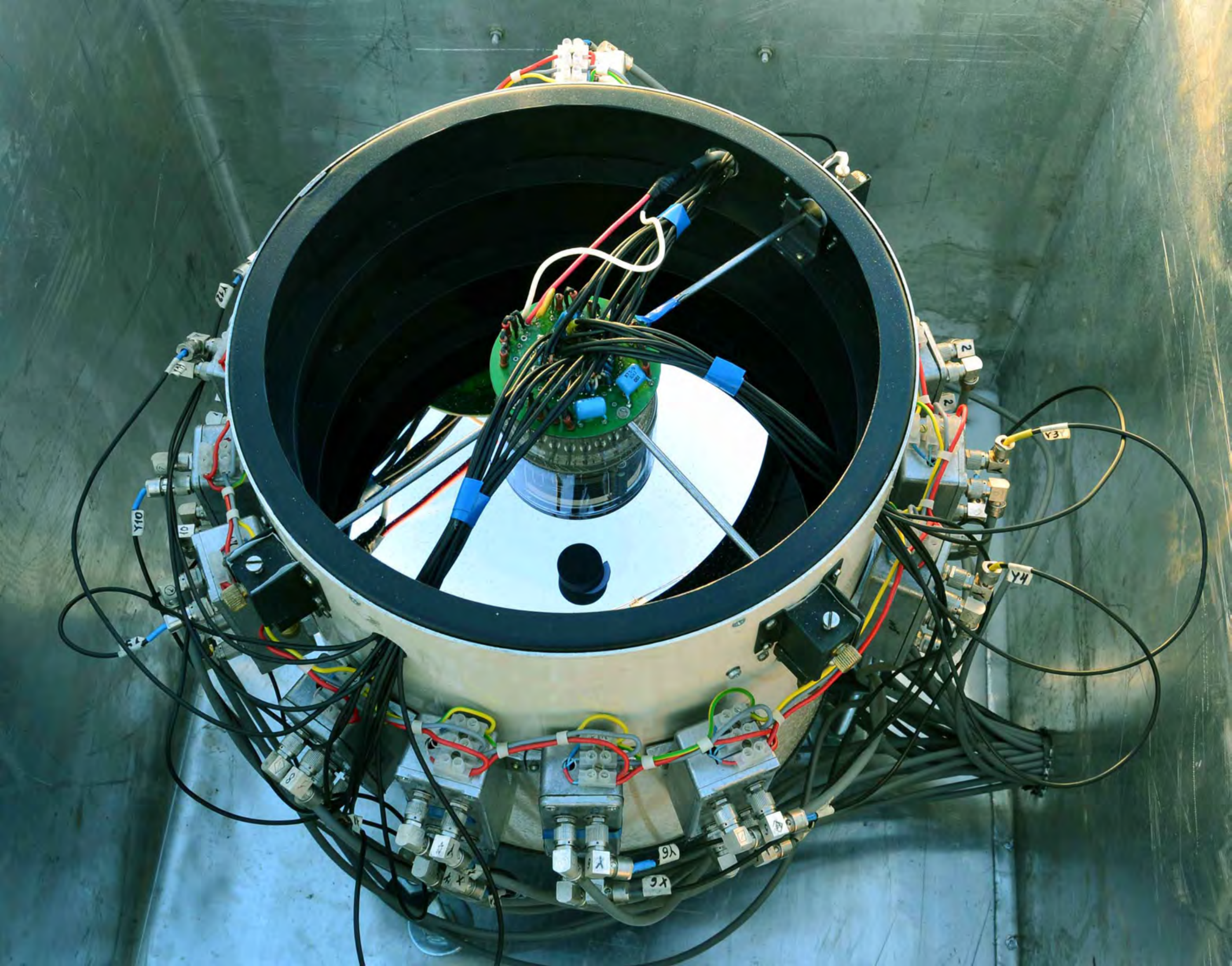}}
\end{minipage}
\caption{The map of surface stations of the Yakutsk array, with distances in meters. Circles indicate stations with scintillation counters producing a trigger; triangles show the central cluster of Cherenkov radiation detectors producing an additional trigger. A cross marks the position of the telescope shown at bottom right. A scintillation counter is shown at the upper right.}
\label{Fig:Array}\end{figure}

The geographical coordinates of the Yakutsk array are ($61.7^{\circ}N,129.4^{\circ}E$) and the site is 100 m above sea level~\cite{MSU,Zenith}. A schematic view of the present layout of the surface stations of the array and photos of a scintillation counter and the telescope are shown in Fig.~\ref{Fig:Array}. Forty-nine stations are distributed within a triangular grid of total area 8.2 km$^2$. The shower events are selected based on coincidence signals from $n\geq 3$ stations, which in turn have been triggered by the two scintillation counters in each station. Complementary triggers at lower energies are produced by the central cluster consisting of 20 Cherenkov radiation detectors~\cite{KnurCERN,KnurFlor,KnurWeihai}.

The main components of the EAS are detected using scintillators, four muon detectors, 48 air Cherenkov light detectors, and six radio detectors.  In this paper, we focus exclusively on the pulse shape of the Cherenkov radiation signal from EAS. Residual aspects concerning other components of the phenomenon are covered in previous papers of the Yakutsk array group~\cite{JETP2007,Tokyo,EMcomponent,FlorenceMass,MinWidth}.

All detectors/controllers and data processing units of the array are connected by a fiber-optic network. An array modernization program is targeted to achieve a LAN channel capacity of 1 Gbps, synchronization accuracy of detectors, and a time resolution accuracy of $10$ ns.
The planned energy range for EAS detection is $(10^{15},10^{19})$ eV~\cite{ASTRA,MainResults}.

\subsection{Wide field-of-view Cherenkov telescope}
\label{sec:Telescope}
The constituent parts of the telescope are a) the spherical mirror (${\o}260$ mm, $f=113$ mm) mounted at the bottom of a metal tube; b) a position-sensitive PMT (Hamamatsu R2486; ${\o}50$ mm) at the focus for which the anode is formed by $16\times16$ crossed wires; c) a voltage-divider circuit and mechanical support attached to the bearing plate; and d) 32 operational amplifiers mounted onto the tube. The telescope is mounted vertically near an array station (Fig. \ref{Fig:Array}).
A comprehensive description of the telescope can be found in Refs. \cite{ASTRA,Tlscp,Tmprl}.

\begin{figure}[t]\centering
\includegraphics[width=0.98\textwidth]{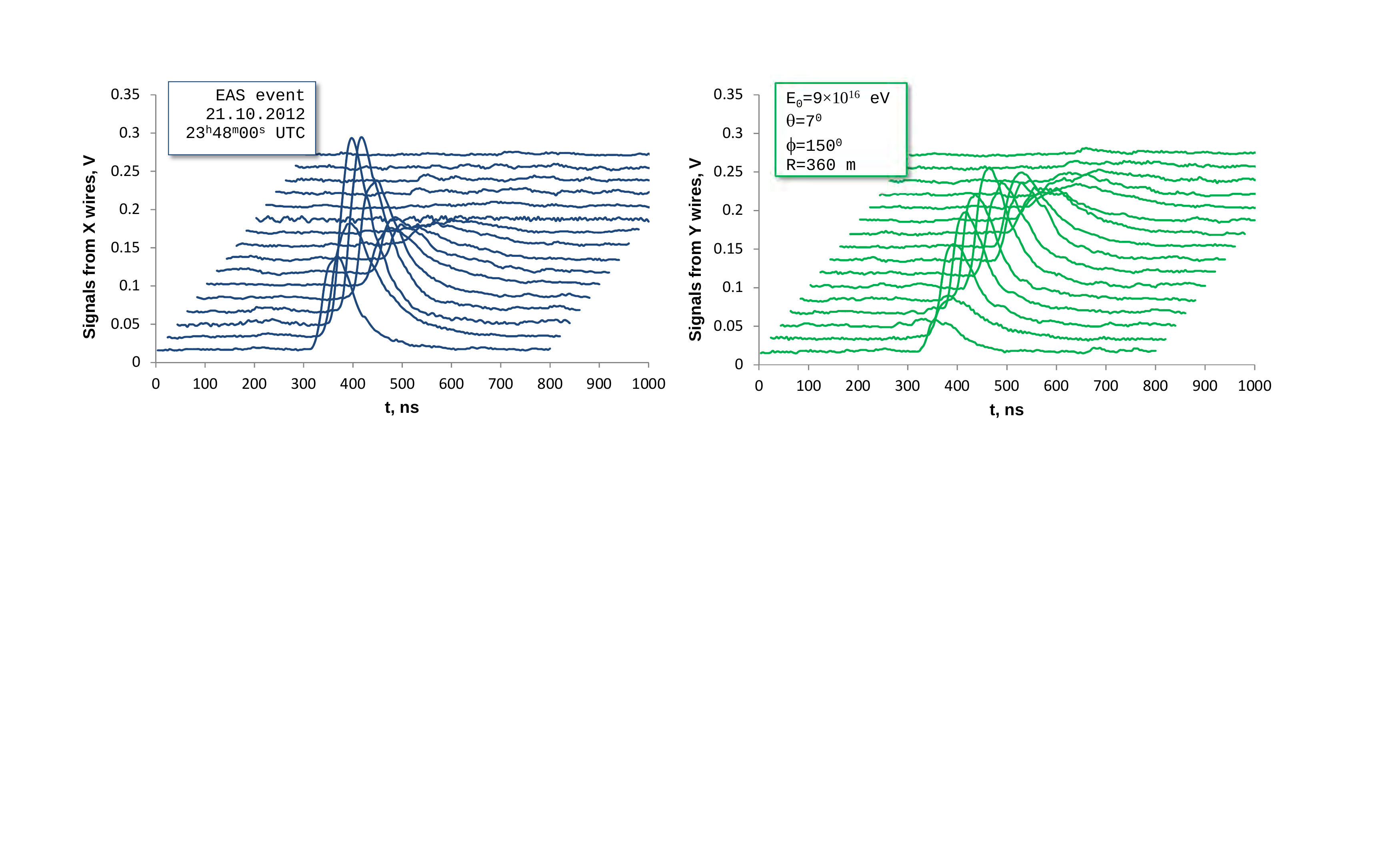}
  \caption{Output signals of WFOV Cherenkov telescope's data acquisition system from an EAS event. Left: signals from X-wires of multi-anode PMT; Right: signals from Y-wires.}
\label{Fig:EASsignal}\end{figure}

The data acquisition system (DAQ) of the telescope consists of 32 operational amplifiers that have 300-MHz bandwidth AD8055 chips connected by long (12 m) coaxial cables to 8-bit LA-n4USB ADC digitizers with 4-ns time slicing. All of the ADC output signals from the 32 channels are continuously stored in PC memory. A trigger signal from the EAS array terminates the process and signals in a 32 $\mu$s interval preceding a trigger are dumped. In Fig. \ref{Fig:EASsignal}, an example is given of the output signals of the DAQ recorded in coincidence with the Yakutsk array detectors in a particular CR shower. The EAS parameters are estimated using the data from the surface array detectors. Nineteen wires exhibit Cherenkov radiation signals; the other thirteen wires do not show a significant signal above the noise level.

In this paper, we are using data accumulated during the period from October 2012 to April 2013
for which EAS events were detected simultaneously by the surface detectors and the telescope.
The other part of the data collected during 2013--2018 is planned to be analyzed. Data selection cuts are applied to exclude showers with cores out of the array area and with zenith angles $\theta>60^{\circ}$. In the present analysis, we are not using the angular dependence of the telescope signals in an individual EAS event; the angular and arrival time differences of signals are ignored. The number of EAS events surviving after cuts is 158.

\section{Deconvolution of Cherenkov Telescope Signals}
\label{sec:Deconvolution}
All of the components of the telescope's DAQ -- amplifiers, ADCs, etc.-- are linear devices; the determined area of linearity for an input signal is (0--0.25) V. The output signal can be represented as a convolution of the input signal $f_{in}(t)$ with a system transfer function $g(t)$

\begin{equation}
f_{out}(t)=\int_{-\infty}^{\infty} f_{in}(\tau)g(t-\tau)d\tau =(f_{in}*g).
\label{Eq:Cnvltn}
\end{equation}

Applying a delta function as input signal, the convolution represents the impulse response of the system. Here, the impulse is a signal composed of all zeros except for a single nonzero point: a digital equivalent of the delta function.

A straightforward way for reconstructing the input signal is a deconvolution using the Fourier transform of signals
\begin{equation}
\hat{f}(\nu)=\int_{-\infty}^{\infty} e^{-2\pi i\nu t}f(t)dt.
\label{Eq:Fourier}
\end{equation}
In the frequency domain, the convolution relation is $\hat{f_{out}}=\hat{f_{in}}\times\hat{g}$ and the deconvolution is given by $\hat{f_{in}}=\hat{f_{out}}/\hat{g}$, where hats symbolize Fourier transforms.

\subsection{Impulse response of the data acquisition system}
\label{sec:Response}
The impulse response of the DAQ -- output of DAQ when presented with a brief input signal (optimum: delta function) -- is ultimately used as a function that is deconvolved from recorded signals in order to obtain unknown input signals. We have tested the DAQ with a signal source that is found brief enough in duration to be considered as a good approximation to a delta function: the dark current impulse of the PMT.
In what follows, durations of all measured signals as a function of time are defined as the full width at half-maximum (FWHM).

A measurement uses the dark current impulse of the shielded PMT produced by an electron beam emitted upon overvoltage. It is important to maintain the anode voltage at a minimal excess of its voltage limit in order to generate a single impulse rather than a multitude of impulses forming the current. The initial pulse shape is determined by the particular PMT characteristics. Our Hamamatsu R2486 position sensitive PMT has a rise time of 5.5 ns.

\begin{figure}[t]\centering
\includegraphics[width=0.7\textwidth]{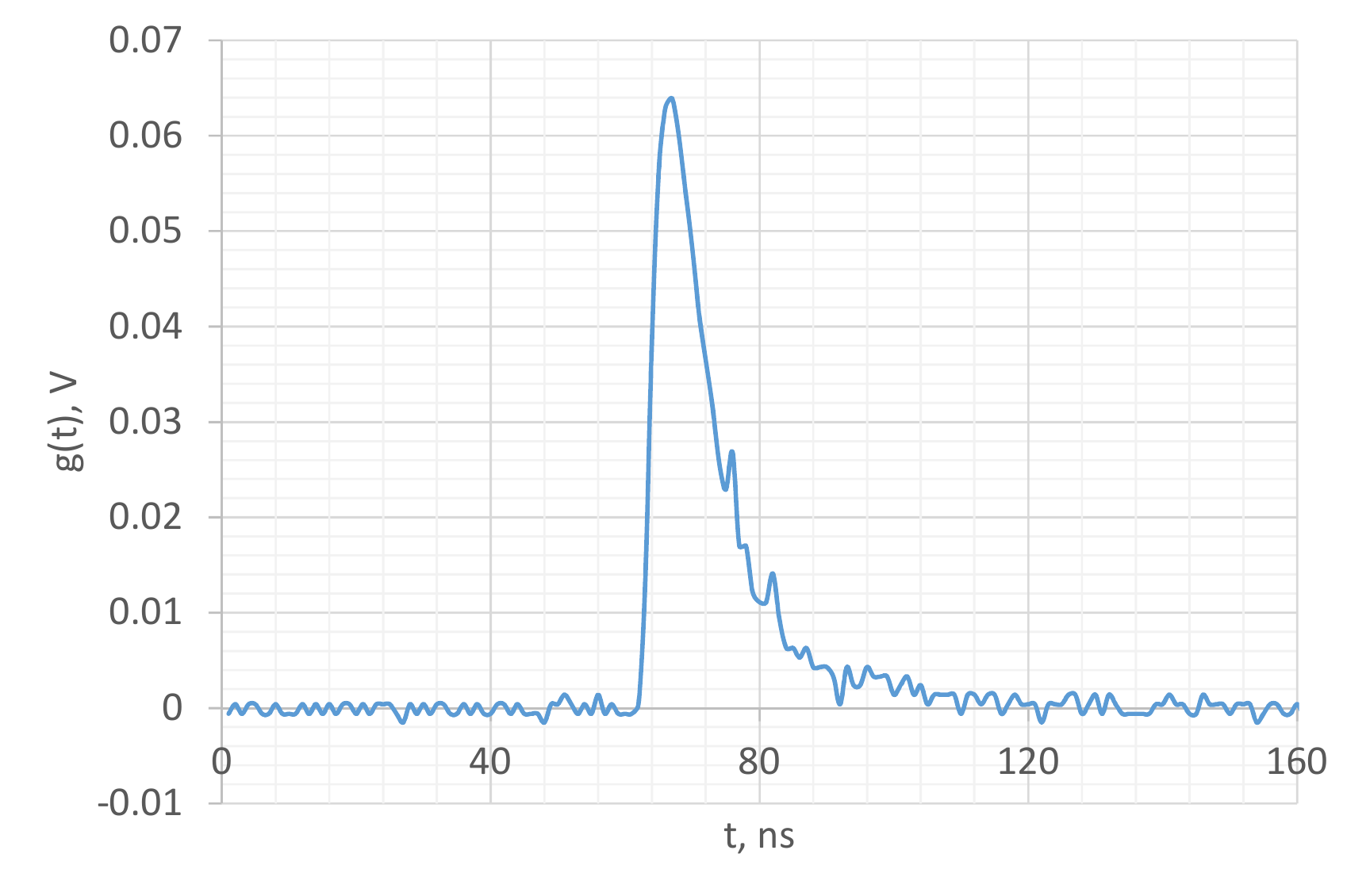}
  \caption{Impulse response of the system to short input signal.}
\label{Fig:Response}\end{figure}

We have measured the duration of the PMT dark current impulse, and found it to be 10.3 ns. The overall duration of the DAQ response is formed cumulatively by amplifiers, ADCs and long lines in the same configuration as for EAS detection.
In Fig.~\ref{Fig:Response}, an example of the resulting impulse response of the DAQ to the dark current impulse is shown.
While a delta function would be ideal, the duration of the output signal determined by the sum of variances is sufficiently longer than that of the input dark current impulse (almost factor $\sim4$). Consequently, the last can be considered as a suitable substitute for the input of a delta function in our case.

The uncertainty of the response, pulse duration in particular, is estimated with a sample of 10 dark impulses. It is found that the mean duration is $36.8\pm 8.8$ ns. The main contribution to the variance, 90\%, is due to dispersion of characteristics of 32 channels, and only 10\% is caused by intrinsic fluctuations of input signals.

\begin{figure}[b]\centering
\includegraphics[width=0.45\textwidth]{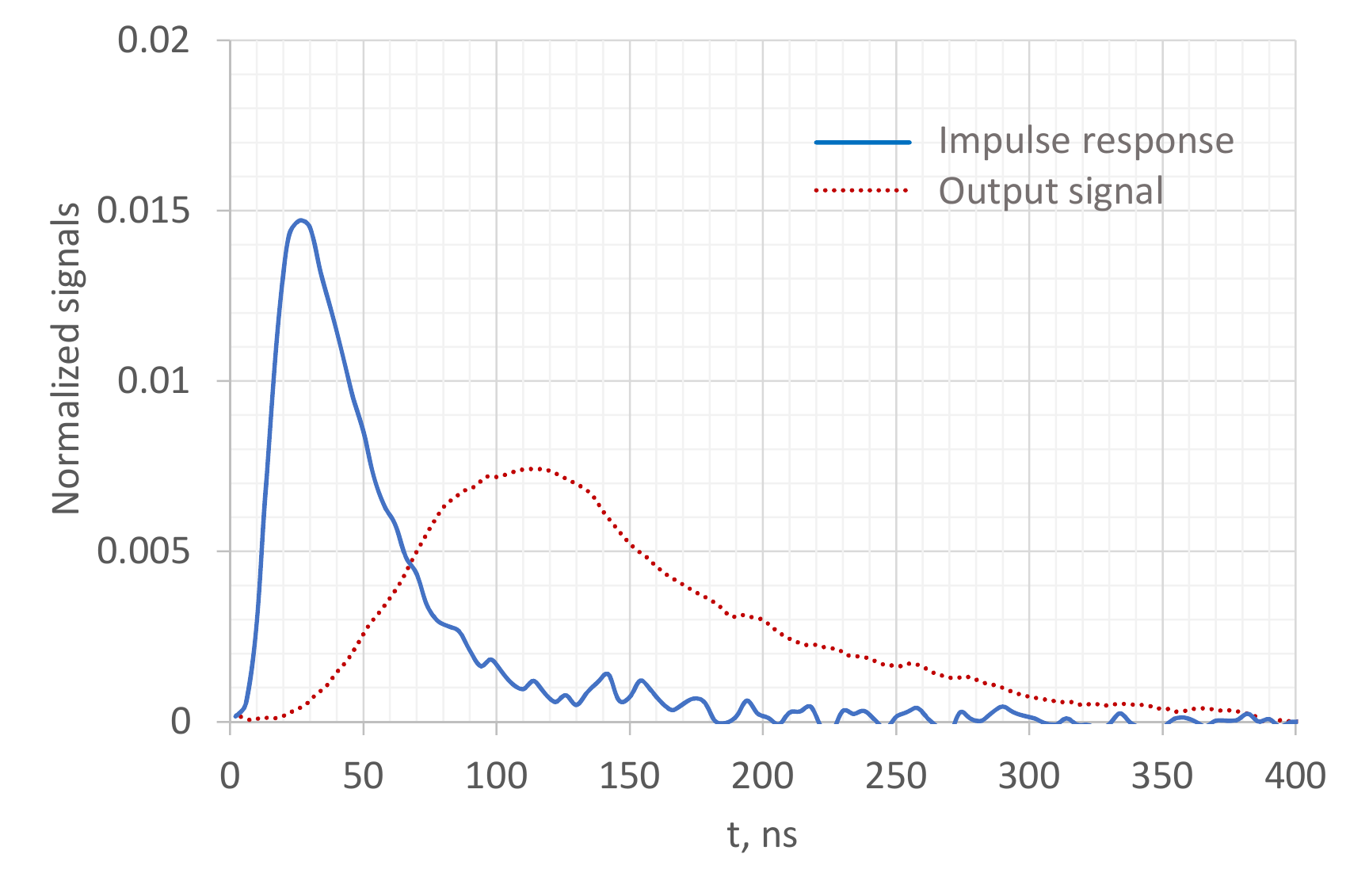}
\includegraphics[width=0.45\textwidth]{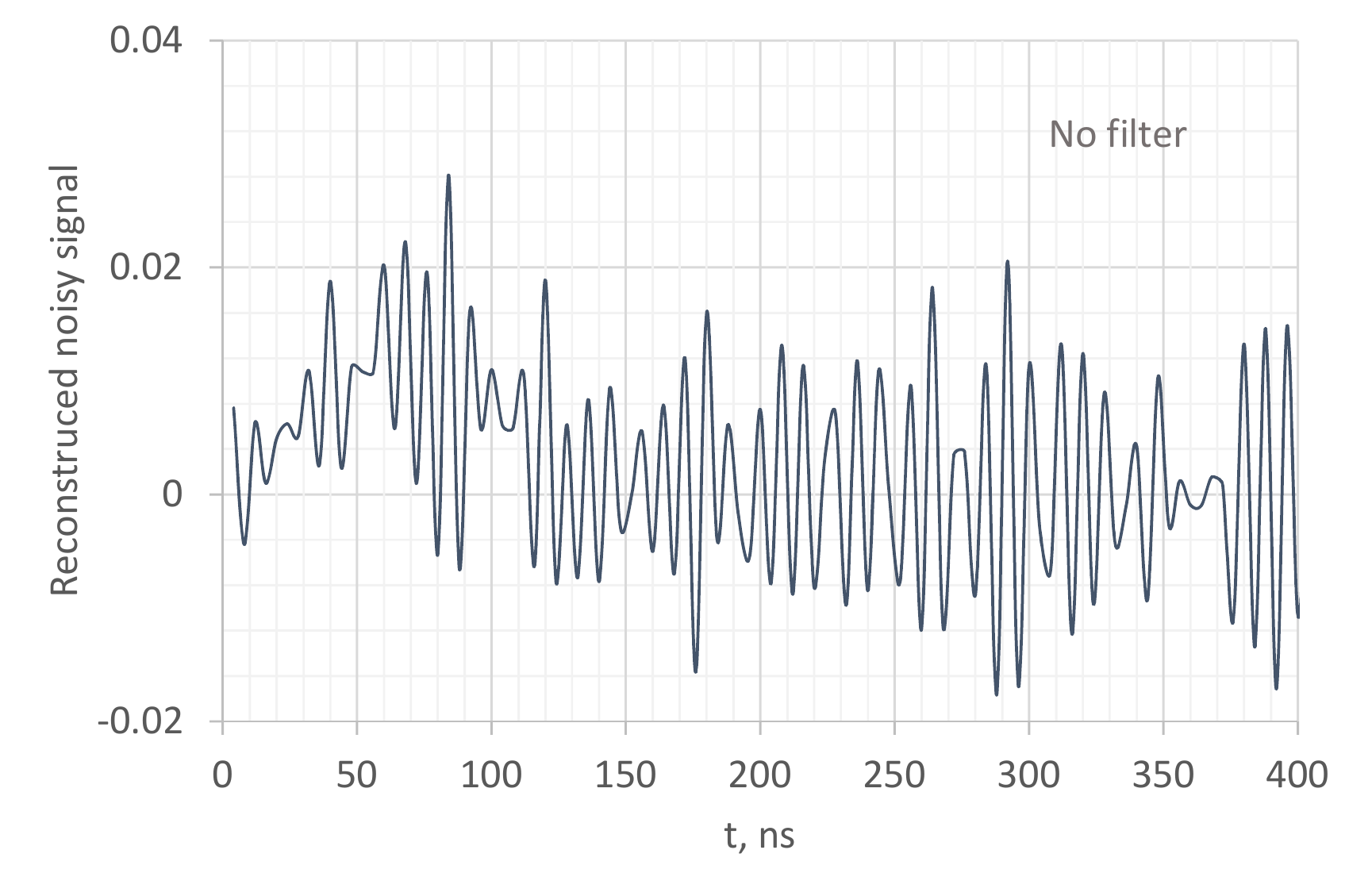}
  \caption{An example of deconvolution of the EAS noisy signal. Left panel: Impulse response and the measured output signal of the system; Right panel: deconvolved input signal.}
\label{Fig:NoFilter}\end{figure}

\subsection{Noise reduction}
\label{sec:Noise}
Having measured the impulse response, we may deconvolve input signals that have durations greater than the duration of $g(t)$. In practice, noise of natural and instrumental origin is added to signals. Comparing the telescope background signals measured with and without a light-proof lid, we have found the fraction of the noise from the night sky light background to be ($72\pm 2$)\%~\cite{Tlscp}.


In Fig.~\ref{Fig:NoFilter}, an illustration is given of a real, noisy input signal which is the result of naive deconvolution of the average measured signal of an EAS (Fig.~\ref{Fig:EASsignal}), using the impulse response presented in the previous section.

In order to carry out the Fourier (and inverse) transforms of digital signals, the fast Fourier transform (FFT) algorithm is used~\cite{FFT}. In the present analysis, we disregard inter-signal time differences in the data from the telescope, shifting signals to the same starting point. The tails of the signals are truncated by an appropriate time window.

From the results shown in Fig.~\ref{Fig:NoFilter}, it is evident that there is a need to use a noise reduction filter for the data from the telescope. We have chosen from the variety of available filters in the field the Wiener filter~\cite{Wnr} working in the frequency domain, characterized by the minimum impact of deconvolved noise at frequencies that have a poor signal-to-noise ratio (S2NR).

\subsubsection{Wiener deconvolution algorithm}
\label{sec:Wiener}
The goal of the approach in this section is to find some function $w(t)$ so that the transform of the input signal can be evaluated as $\hat{f_{in}}=\hat{w}\hat{f_{out}}$, where an estimate of $f_{in}$ minimizes the mean square error. Wiener showed that the filter
\begin{equation}
\hat{w}=\frac{1}{\hat{g}}[\frac{|\hat{g}|^2}{|\hat{g}|^2+\frac{N(f)}{S(f)}}]
\label{Eq:Wnr}
\end{equation}
provides such a function in the frequency domain, where $N(f)$ and $S(f)$ are the mean power spectral density of the noise and input signal, respectively~\cite{Wnr}. When the noise is zero, the expression reduces to $\hat{f_{in}}=\hat{f_{out}}/\hat{g}$. The ratio $S(f)/N(f)$ is the representation of the S2NR. Therefore, the Wiener filter attenuates frequencies depending on the signal-to-noise ratio.

The time window of the DAQ of the telescope is 32 $\mu s$, while the maximum duration of the Cherenkov signal is 300 ns,
so we have more than hundred time windows
where only the noise is detected from which $N(f)$ can be estimated. On the other hand, we do not know the power spectral density of the input signal. The only possibility is to estimate it using the output signal.

\begin{figure}[b]\centering
\includegraphics[width=0.7\textwidth]{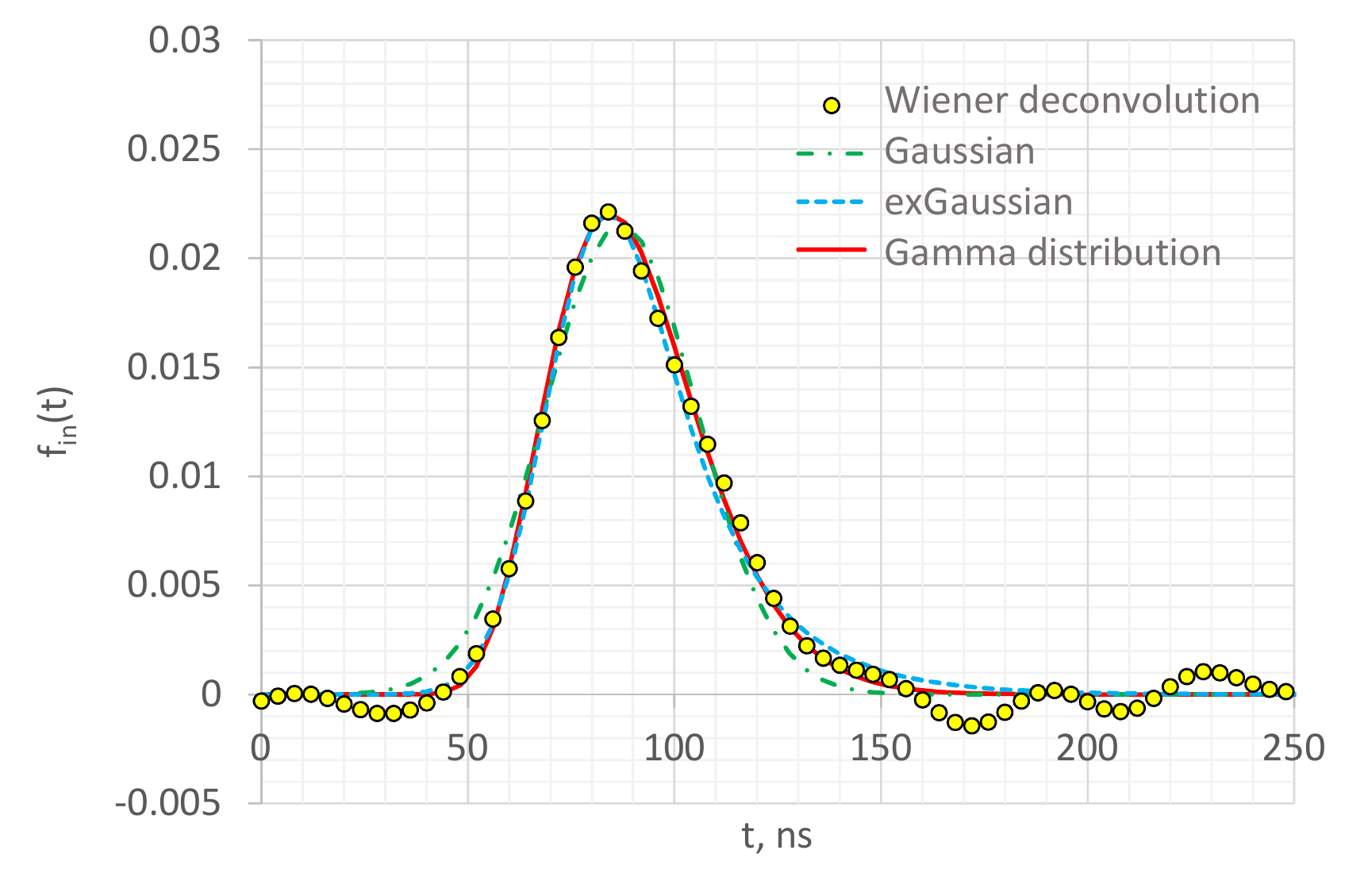}
  \caption{Analytic approximations to the deconvolved input Cherenkov signal from EAS.}
\label{Fig:Fits}\end{figure}

In Fig.~\ref{Fig:Fits} a sample of the resulting Wiener deconvolution of signals of the EAS event detected on 21.10.2012 (Fig.~\ref{Fig:EASsignal}) is shown. Time windows of equal width are set for input signals and noise. Fourier transforms of the signals and power spectral densities are derived using an FFT program in Fortran (code 12-4 in the book by Smith~\cite{DSP}).

\subsubsection{Gamma distribution as an approximation to the input signal}
\label{sec:GD}
It seems that the reconstructed input signal of the Cherenkov radiation induced by EAS can be approximated by some kind of common functions. We have tested the most suitable distributions as an analytic approximation to the input Cherenkov signal from the EAS event detected on 21.10.2012. As a goodness-of-fit criterion the minimum sum of square deviations from experimental points is used.

Three probability distributions tested in our analysis are: a) Gaussian
\begin{equation}
f(x)=\frac{1}{\sqrt{2\pi D}}\exp(-\frac{(x-\mu)^2}{2D}),
\label{Eq:Gauss}\end{equation}
where $\mu$ is the mean value, and D is the variance;
b) exponentially modified Gaussian (exGaussian)
\begin{equation}
f(x)=\frac{\lambda}{2}\exp(\frac{\lambda}{2}(2\mu+\lambda D-2x))erfc(\frac{\mu+\lambda D-x}{\sqrt{2D}}),
\label{Eq:Gauss}\end{equation}
where $\mu$ and $D$ are the same as in the Gaussian distribution, and $\lambda$ is the rate;
c) Gamma distribution
\begin{equation}
f_\gamma(x)=\frac{1}{\Gamma(\kappa)\tau^\kappa}(x)^{\kappa-1}e^{-x/\tau},
\label{Eq:GD}\end{equation}
where $\kappa$ and $\tau$ are shape and scale parameters, and $\Gamma$ is the normalizing gamma function. The mean value and variance are given by $\kappa\tau$ and $\kappa\tau^2$, respectively.

The gamma distribution having the minimum sum turns out to be the best approximation to the reconstructed Cherenkov signal (the sum is $2.4\times 10^{-5}$ against $6.9\times 10^{-5}$ for Gaussian and $4.0\times 10^{-5}$ for exGaussian). A comparison of signal durations in 12 channels with the largest S2NRs of the EAS event shown in Fig.~\ref{Fig:EASsignal}  is given in the first two rows of Table 1. A difference in FWHM of the deconvolved and fitted gamma distributions is less than $10$ \%. Comparison of the shape of deconvolved input signal and three approximations is given in Fig.~\ref{Fig:Fits}.

\begin{table}[t]
\tbl{Duration of input signals, FWHM, ns.}
{\begin{tabular}{lrrrrrrrrrrrr}
\hline\hline
Channel number         &  1   &  2   &  3   &  4   &  5   &  6   &  7   &  8   &   9  &  10  &   11 &   12 \\ \hline
Wiener deconvolution   & 41.6 & 51.7 & 37.3 & 35.6 & 42.5 & 32.1 & 56.1 & 45.7 & 45.1 & 40.0 & 47.5 & 58.0 \\
Gamma distribution fit & 41.5 & 52.0 & 38.1 & 36.6 & 42.4 & 35.2 & 55.8 & 45.2 & 47.3 & 39.2 & 47.8 & 58.0 \\
A simple algorithm     & 26.2 & 52.3 & 28.6 & 27.7 & 29.8 & 30.1 & 66.7 & 51.8 & 54.2 & 29.3 & 33.5 & 37.6 \\
Ratio (Wiener/Simple)  & 1.59 & 0.99 & 1.31 & 1.28 & 1.43 & 1.07 & 0.84 & 0.88 & 0.83 & 1.36 & 1.42 & 1.54 \\
\hline\hline
\end{tabular}}
\end{table}

\subsubsection{A toy model}
\label{sec:Toy}
We introduce a simple model in this section in order to illustrate the reconstruction of the input signal and in order to elucidate the influence of noise on the Wiener deconvolution result. In accordance with the conclusion in the previous section, we use a gamma distribution as the input signal and the real impulse response given in Section~\ref{sec:Response}.

\begin{figure}[t]\centering
\includegraphics[width=0.7\textwidth]{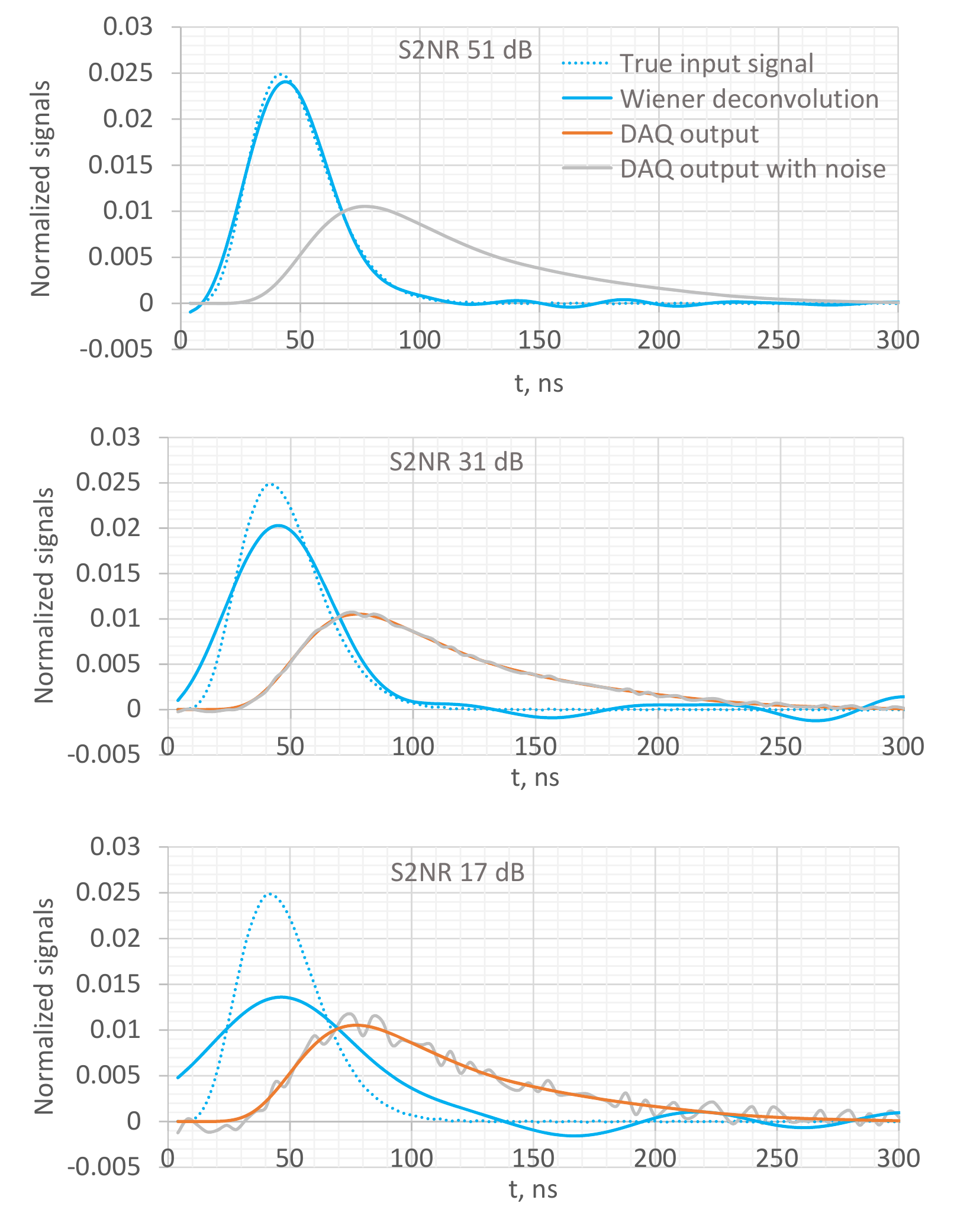}
  \caption{Wiener deconvolution of a noisy signal in a toy model.}
\label{Fig:Toy}\end{figure}

A background of white noise (a sequence of uncorrelated signals) is modeled using a pseudo-random number generator in Fortran $x_i=2\times A_{noise}\times(ran(i)-0.5)$, where $A_{noise}$ is a constant noise amplitude: $x_i\in(-A_{noise},A_{noise})$. The noise is added to $f_{out}$, which is formed by the convolution $(f_{in}*g)$. The signal-to-noise ratio is defined by the variances of signal, $\sigma^2_{signal}$, and noise, $\sigma^2_{noise}$, expressed in decibels: S2NR=$10\lg(\sigma^2_{signal}/\sigma^2_{noise})$. The ratio is calculated in the same time window for signal and noise.

\begin{figure}[t]\centering
\includegraphics[width=0.7\textwidth]{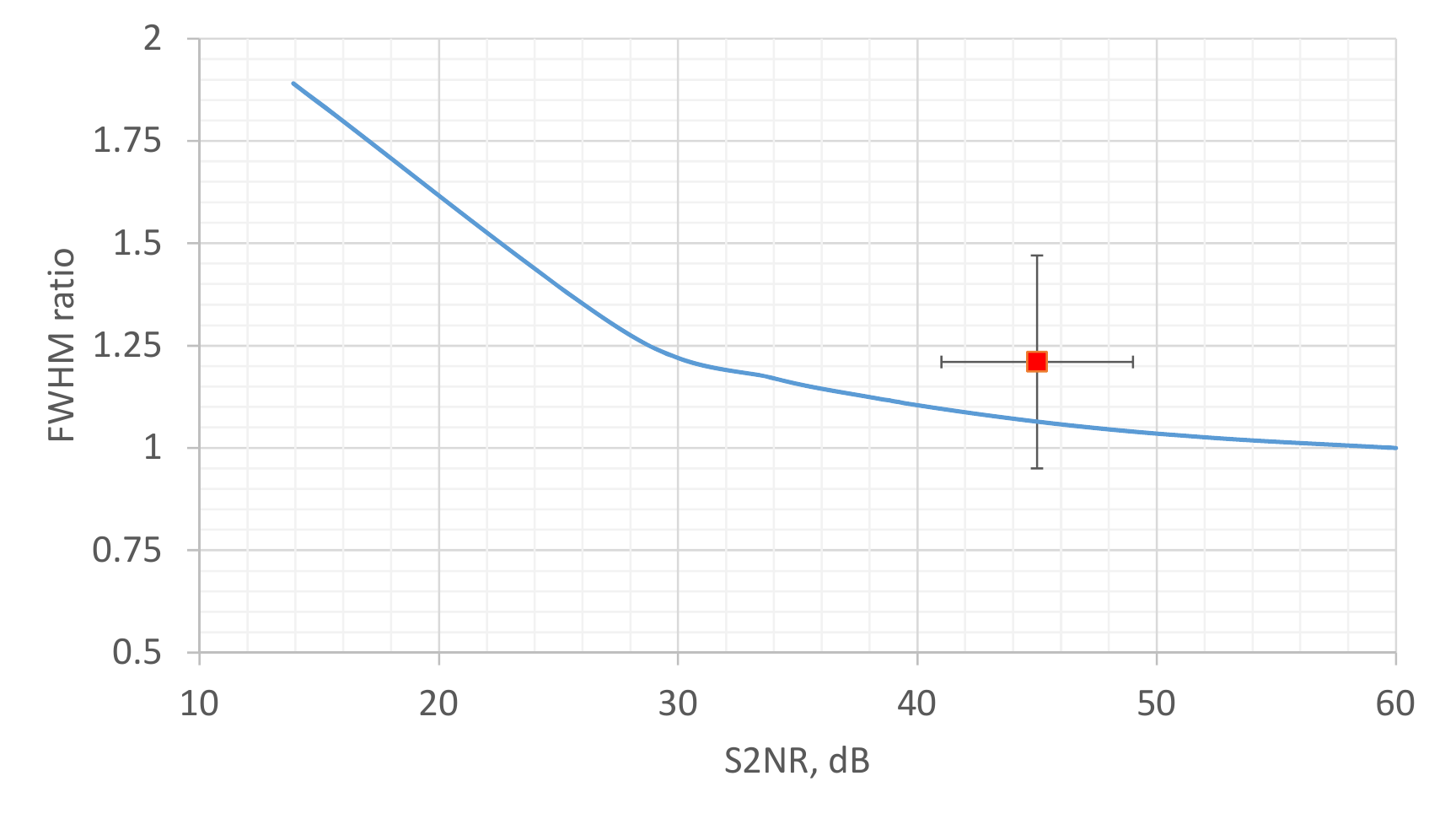}
  \caption{The ratio of durations of Wiener deconvolution and input signals vs. S2NR in a toy model (solid curve). The red square indicates the ratio of estimated signal durations of the telescope, as described in Section~\ref{sec:Ginput}}.
\label{Fig:ToyDur}\end{figure}

The input signal is reconstructed using the Wiener deconvolution as described in Section~\ref{sec:Wiener}. The resulting signal is compared with the true input in Fig.~\ref{Fig:Toy} for different S2NR. Increasing the noise fraction leads to a distortion of the reconstructed signal. The necessary condition on S2NR in order to have an artifact in the Wiener deconvolution less than 1\% is S2NR$>56$ dB. In Fig.~\ref{Fig:ToyDur}, the ratio of FWHM of reconstructed and true input signals as a function of S2NR is shown by the solid curve.

\section{Analysis of Temporal Characteristics of Cherenkov Radiation Signals Detected by the Telescope in Coincidence with the EAS array}
\label{sec:Data}
The telescope detects EAS in two modes: independently and in coincidence of signals with the surface array detectors. In the present work, only showers detected in the second mode are analyzed. Coincident EAS events are selected via trigger signals from the array generated when three or more detectors are hit by a shower. The total number of triggers accumulated during the period October 2012--April 2013 is 29296, while the number of coincident EAS events detected by the array and telescope is 701. The ratio of events conforms to the ratio of the array area and acceptance area of the telescope~\cite{Tlscp}. After quality cuts the number of coincident events under analysis reduces to 158. The mean energy of the primary particles initiating EAS in a sample is $(2.5\pm0.3)\times10^{17}$ eV, and the mean zenith angle of showers is $20^{\circ}\pm3^{\circ}$.

\subsection{Gamma distribution convolution method of signal reconstruction}
\label{sec:Ginput}
The finding in Section~\ref{sec:GD} that the input Cherenkov radiation signal from EAS can be approximated by a scaled gamma distribution function gives a possibility to implement a simplified, fast signal reconstruction algorithm omitting the deconvolution step (`forward folding'). Namely, the method of gamma distribution convolution (GDCM) consists of adjustment of the time window to $f_{out}$ and fitting the free parameters of $f_{in}^{\gamma}$ so that the convolution result is congruent to the measured output signal. In that case the representation of $f^{\gamma}_{in}$ in the form of Eq.~(\ref{Eq:GD}) merely leaves two parameters ($\tau,k$) to modify the output signal $f_{out}=(f_{in}^{\gamma}*g)$. In practice, two other related parameters of the distribution are convenient: the mean value, $\mu=\tau\kappa$,  and r.m.s. deviation $\sigma=\tau\sqrt{\kappa}$.

\begin{figure}[t]\centering
\includegraphics[width=0.7\textwidth]{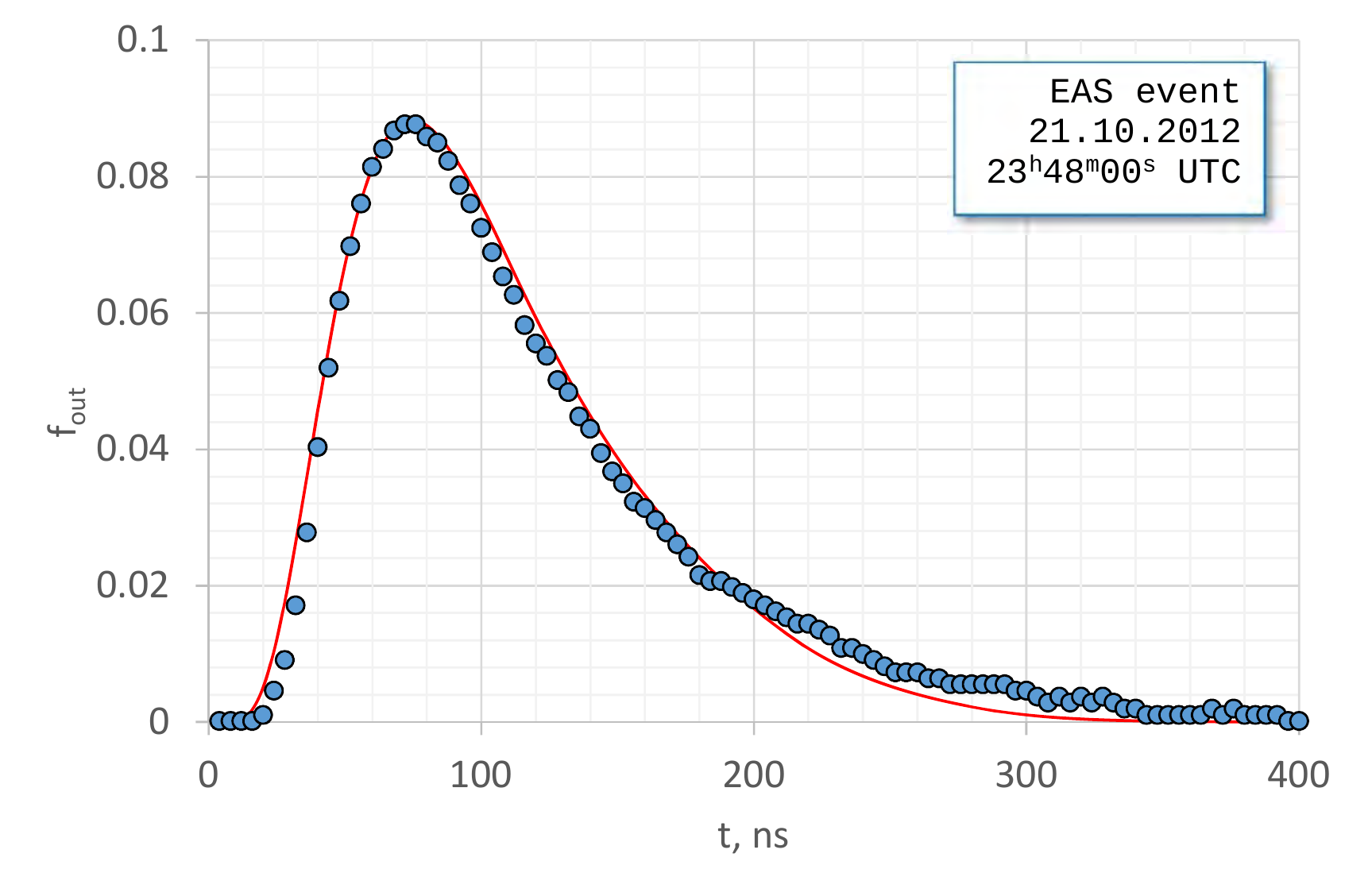}
\caption{Measured vs. modeled output signal with optimized gamma distribution as input signal.
Horizontal axes: t, ns; vertical axes: signals in arbitrary units. Experimental data are indicated by points. The convolution of the impulse response and trial gamma distribution as an input signal is shown by the solid curve.}
\label{Fig:Simple}\end{figure}

To evaluate the parameters of the gamma distribution, the non-linear least squares approach is used. The aim is to minimize the sum of the squares of differences between the observed signal and the convolution result in the time window. The optimal values of parameters are found applying a numerical Gauss--Newton algorithm~\cite{GaussNewton}. An example of a fitted output signal of the particular EAS event is shown in Fig. \ref{Fig:Simple} in comparison with experimental data. It appears that by having two adaptable parameters of the input gamma distribution, we may obtain a sufficient description of the observed output signal. Another application of the method is to recover saturated signals in the case of measuring Cherenkov radiation from EAS (see~\ref{sec:Saturated}).

\subsection{A search for correlation between parameters of gamma distribution and showers}
\label{sec:Correlation}
As was shown with the toy model, the additive noise influence leads to a distortion of the deconvolved input signal, increasing with increased noise fraction, at least for the Wiener algorithm. In order to estimate the accuracy of the reconstructed signal, two independent methods may be compared: a) the Wiener deconvolution and b) a GDCM reconstruction of the input signal in the form of a gamma distribution. A comparison of durations of the input signals is presented in the final two rows of Table 1. The ratio of FWHM of Wiener deconvolution results and the gamma distribution width is, on average, $1.21\pm0.26$. This result may be attributed to S2NR=($45\pm4$) dB\footnote{Estimated as S2NR=$10\lg((\sigma^2_{Fout}-\sigma^2_{noise})/\sigma^2_{noise})$ in 12 channels} inherent in measurements of the telescope of the Yakutsk array. The ratio of durations of the experimental signals is compared with the results of the toy model in Fig.~\ref{Fig:ToyDur}.

To decrease the influence of noise on the analyzed signals, we selected DAQ output signals with amplitudes above the threshold 0.075 V. Saturated signals are omitted, too. Gamma distribution parameters are averaged for every EAS event over channels surviving after cuts. For instance, in the event no. 034 detected 21.10.2012, $23^h48^m00^s$ UTC (Fig.~\ref{Fig:EASsignal}) only twelve channels produced signals with amplitudes $>0.075$ V.

Having characteristics of showers estimated based on EAS array data and parameters of the gamma distribution which approximate the input Cherenkov radiation signal, we have calculated Pearson's correlation coefficients, $\rho_{x,y}$, between them using a sample of 158 coincident events
\begin{equation}
\rho_{x,y}=\frac{cov(x,y)}{\sigma_x\sigma_y},
\label{Eq:cov}\end{equation}
where $x,y$ are variables; cov means covariance, and $\sigma_x,\sigma_y$ are standard deviations.

In Table 2, resultant correlation coefficients of the gamma distribution parameters with the main shower characteristics, $E,R,\theta$, are given. No appreciable correlation is found, except FWHM of $f_{in}(t)$ vs. $R$, the core distance. In the previous paper~\cite{Tmprl}, we estimated the duration of the input signal without deconvolution, by means of difference of variances of the output signal and DAQ response.

Now, we can deconvolute the Cherenkov radiation signal applying GDCM in a more reliable way.

\begin{table}[t]\centering
\tbl{Correlation coefficients between gamma distribution parameters (expected value, $\mu$; standard deviation, $\sigma$; full width at half-maximum, FWHM) and EAS characteristics.}
{\begin{tabular}{lrrr}
\hline\hline
$f_\gamma(t)$ parameters & $\mu$ & $\sigma$ &  FWHM \\ \hline
Primary energy, $E$      & 0.004 & -0.022   &  0.163\\
Core distance, $R$       & 0.103 & -0.085   &  0.670\\
Zenith angle, $\theta$   & 0.118 &  0.215   & -0.187\\
\hline\hline
\end{tabular}}
\end{table}

\subsection{Duration of Cherenkov radiation signal as a function of EAS core distance}
\label{sec:FWHM}
Durations of all measured signals are estimated numerically at half-maximum employing linear interpolation of digital signals.
The distance between the detector and the shower core, $R$, is defined as the bee line to the shower axis
\begin{equation}
R=L\sqrt{\sin^2\psi+\cos^2\psi\cos^2\theta},
\label{Eq:Rp}\end{equation}
where $L$ is the distance to the core in the array plane, and $\psi$ is the angle between $L$ and the shower axis projection.

\begin{figure}[b]\centering
\includegraphics[width=0.7\textwidth]{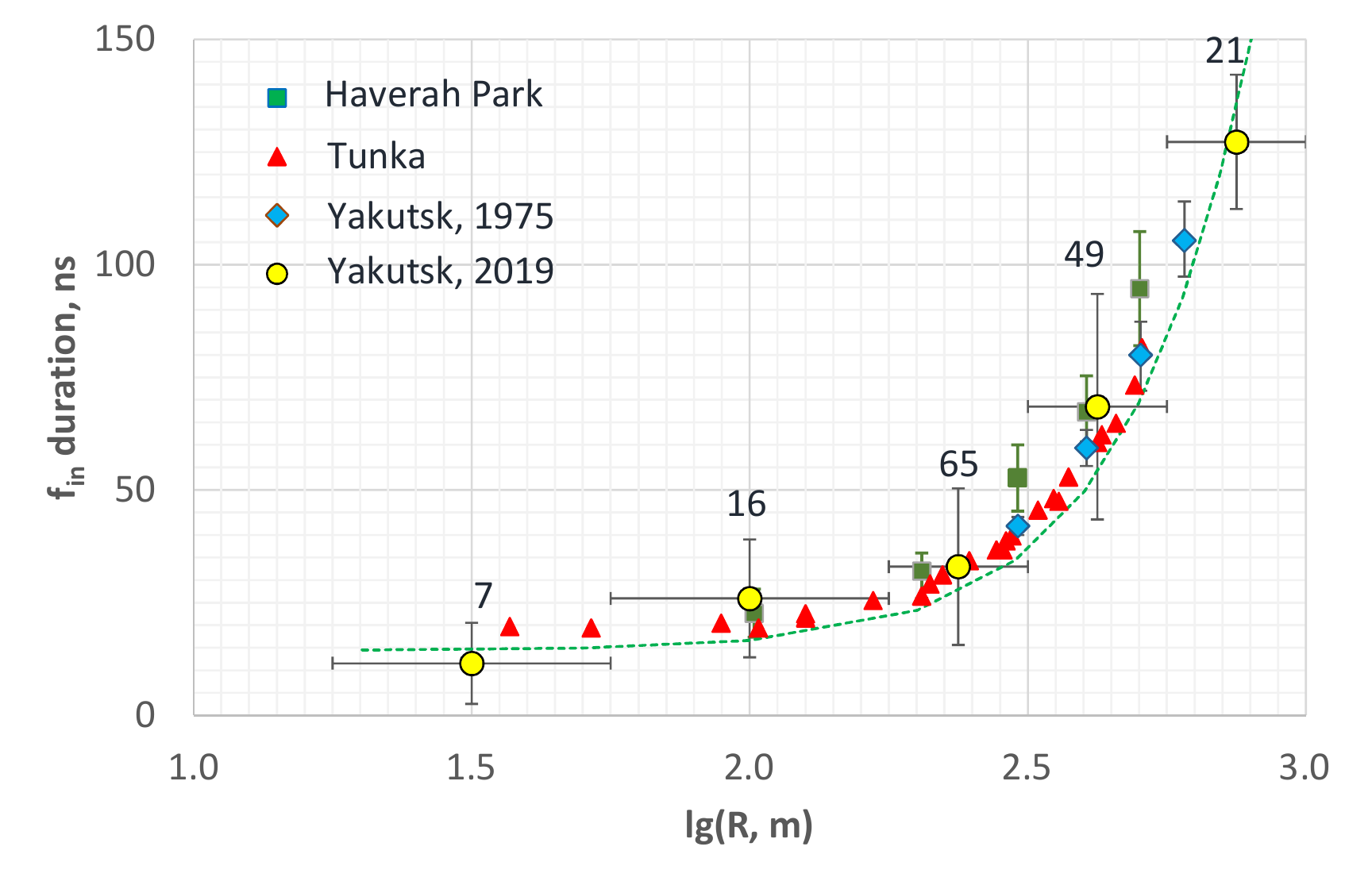}
\caption{FWHM of Cherenkov radiation signals induced by EAS vs. shower core distance. Present results are deconvolved from the data of the Cherenkov telescope working in coincidence with the Yakutsk array detectors. Vertical and horizontal bars indicate experimental errors and $\lg(R)$ interval width. EAS event numbers in the intervals are shown above data points. The model behaviour of the signal duration as a function of $\lg(R)$ is given by the dashed line.}
\label{Fig:Duration}\end{figure}

Previous measurements of the Cherenkov signal duration as a function of the shower core distance were made in Haverah Park~\cite{Turver}, Yakutsk~\cite{Klmkv} and Tunka~\cite{Tunka} arrays.
Data from the Tunka experiment should be regarded as a preliminary estimation of the upper limit to the signal duration, because: 1) FWHM$(R)$ function is obtained without deconvolution of the observed output signal; 2) measurements are made in a single EAS event, therefore, the event-by-event fluctuations of the shower parameters are not considered.
Our evaluation of the FWHM$(R)$ qualitatively confirms previous results and expands the range of core distances involved in a single experiment (Fig. \ref{Fig:Duration}).

A distinctive feature of the Cherenkov signal duration is its causal relationship with the geometric size of the light pool of EAS only. In order to elucidate the magnitude of the signal duration and the core distance dependence, it is sufficient to consider a back-of-the-envelope calculation model of the shower structure.

\subsubsection{A region near the shower core}
It can be concluded from experimental results in Fig. \ref{Fig:Duration} that at distances $R<50$ m the Cherenkov signal duration converges to the minimal limit. Indeed, one can estimate the signal duration at $R=0$ as a time difference between photons arriving to the detector from the array level and the upper atmosphere

\begin{equation}
\Delta t\sim\frac{h_0}{c}((1-\gamma^{-2})^{-0.5}-n),
\label{Eq:Rp}\end{equation}

where $h_0$ is the height in the atmosphere where the radiation begins, $\gamma$ is the Lorentz factor of emitting particles, and $n\sim 1+2.9\times 10^{-4}$ is the refractive index of air. We used a conventional value of the maximum depth measured in HiRes~\cite{HiRes}, Tunka~\cite{TunkaComposition} and Yakustk~\cite{Knur24ECRS} experiments $x_{max}\sim650$ g/cm$^2$ (the corresponding height in Yakutsk is $h_{max}^e\sim3.1$ km) for the number of electrons in EAS with energy $E_0=2.5\times10^{17}$ eV and zenith angle $\theta=20^{\circ}$ as an upper limit to the radiation region's height. As shown previously, the Cherenkov radiation emission maximum in EAS is less than $h_{max}^e$ due to the angular distribution of relativistic electrons \cite{Tlscp}.  The minimum of $\gamma=41.3$ is caused by the Cherenkov radiation condition $v>c/n$ in air. The resultant time difference is below 4 ns. It is significantly less than the instrumental uncertainty of the experiment and the measured durations.

Another source of time differences is the lateral width of the light emitting region with radius $R_{Cher}$ at the height in the atmosphere, $h_{max}^{Cher}$, where the maximum of radiation is located

\begin{equation}
c\Delta t\sim\sqrt{(h_{max}^{Cher}\sec\theta)^2+4R_{Cher}^2}-(h_{max}^{Cher}\sec\theta)^2.
\label{Eq:Dch}\end{equation}

Assuming the upper limit of the measured signal duration at $R<50$ m to be 20 ns and $h_{max}^{Cher}<3.1$ km we conclude that $R_{Cher}$ should be less than 100 m.

\subsubsection{A region far from the core}
At distances from the shower core $R\gg R_{Cher}$, the width of the radiation region can be neglected. Here, the length of the shining area along the shower axis can be estimated instead. The radiation is modeled in a shower as the shining section of the axis with the length $L$ around $h_{max}^{Cher}\sec\theta$. The time difference at the distance $R$ is

\begin{equation}
c\Delta t\sim L+\sqrt{R^2+(h_{max}^{Cher}\sec\theta-0.5L)^2}-\sqrt{R^2+(h_{max}^{Cher}\sec\theta+0.5L)^2}.
\label{Eq:Lch}\end{equation}

Assuming $R=750$ m, $dt=130$ ns, and $h_{max}^{Cher}<3.1$ km, we set an upper limit to the length of the shining area $L<1500$ m. It turned out that the Cherenkov radiation emitting region in EAS, distinguished by the intensity above a half of the maximum, is of cylindrical form with diameter and length less than 200 and 1500 m, correspondingly.

Furthermore, our model can describe the shower core distance dependence of the signal duration as well. It is convenient to insert into Eq.~(\ref{Eq:Lch}) an estimation of the width of the shining area as in Eq.~(\ref{Eq:Dch}) and then fit the parameters. The resultant dependence of FWHM$(R)$ is given in Fig.~\ref{Fig:Duration} by the dashed line.

\section{Conclusions}
We have applied digital signal processing in order to reconstruct Cherenkov radiation signals, induced by EAS, from the data of a telescope working in co-operation with the surface scintillation counters of the Yakutsk array.

The transfer function of the data acquisition system is evaluated using a dark current impulse of the multi-anode PMT. Using this transfer function and a Wiener deconvolution algorithm, the input signal is reconstructed. The influence of noise parameterized with the signal-to-noise ratio is estimated using a toy model.

It is demonstrated that the Cherenkov radiation signals from EAS can be approximated by a scaled gamma distribution. Consequently, an efficient model-independent method for reconstruction of such signals is proposed that does not require a simulation of the shower development in the atmosphere. As an additional bonus, the reconstruction method can be used to recover saturated output signals.

A dataset of coincident EAS events detected by the telescope and surface array detectors during the period October 2012--April 2013 is analyzed. A significant correlation of the Cherenkov radiation signal duration with the distance to the shower core is found in agreement with previous measurements. The results obtained in an extended radial interval enabled us to set an upper limit to dimensions of the area along the shower axis where the Cherenkov radiation intensity is above half-peak amplitude.

The length of the shining area is found to be less than 1500 m, and the diameter is less than 200 m in EAS with the primary energy $E_0=2.5\times10^{17}$ eV and zenith angle $\theta=20^{\circ}$, insofar as the height of the maximum Cherenkov radiation in the atmosphere is less than the maximum height of the number of electrons.

\section*{Acknowledgments}
We would like to thank the Yakutsk array group for data acquisition and analysis. This study is supported in part by the Presidium of RAS (program 3), the Siberian Branch of RAS (program II.16.2.3), and RFBR (project 16-29-13019).

\appendix
\section{Reconstruction of Saturated Signals}
\label{sec:Saturated}

Some portions of the detected signals in the experiments are saturated because of the maximum capacity of the instruments having been exceeded. This behavior is commonplace in cosmic ray physics, particularly in EAS studies in which there is a wide dynamic range of energy ($10^{14},10^{20}$) eV.

While a conventional approach to repairing the saturated signal is to avoid it, in some cases, when the most physically interesting information is lost, or the measurement is unique, it is preferable to recover the signal in the primary saturation domain. Several methods for the reconstruction of saturated signals have been proposed previously~\cite{Yang,Liu}.

In our case, the input Cherenkov radiation signal can be approximated using a gamma distribution, offering a simple method for reconstruction of the saturated output signal. The algorithm is the same as in Section~\ref{sec:Ginput}, besides ignoring the cut-off portion of the signal. Obviously, the accuracy of a recovery depends on the fraction of signal that was lost to saturation.

\begin{figure}[t]\centering
\includegraphics[width=0.45\textwidth]{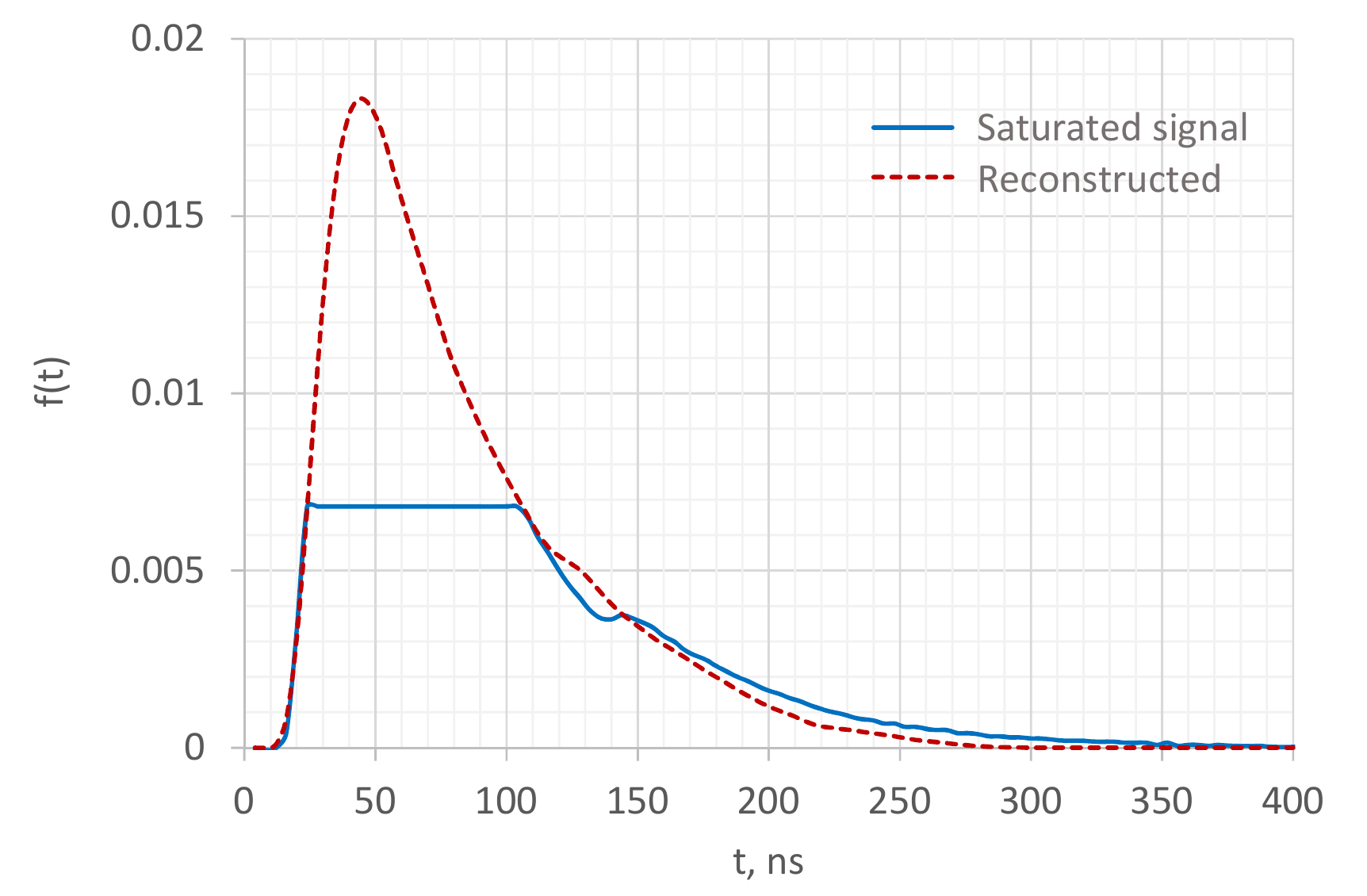}
\includegraphics[width=0.45\textwidth]{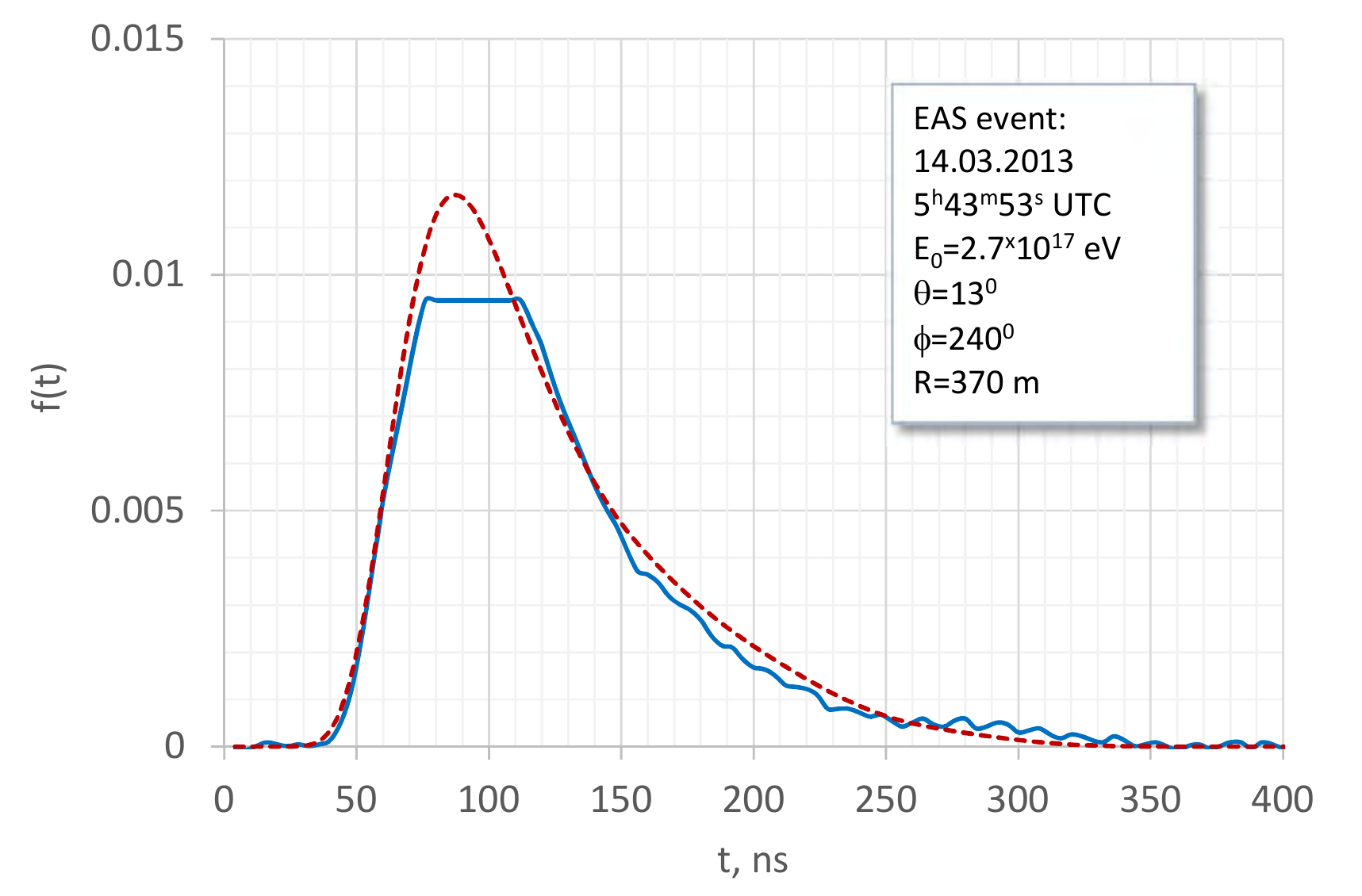}
  \caption{Reconstruction of two saturated output signals from the EAS event denoted in the right panel.}
\label{Fig:Satur}\end{figure}

In Fig. \ref{Fig:Satur}, two signals reconstructed using this method are shown. For an EAS event detected 14.03.2013, $5^h 43^m 53^s$ UTC, signals for two of the 32 wires in the telescope are saturated. In the second case, a relatively insignificant portion of the signal is lost and the reconstruction is more reliable -- at least, in terms of signal duration.

\end{document}